\documentclass[sigconf,natbib=false, anonymous=false]{acmart}

\AtBeginDocument{%
  }

\copyrightyear{2022} 
\acmYear{2022}
\setcopyright{rightsretained}
\acmConference[ICAIF '22]{3rd ACM International Conference on AI in
Finance}{November 2--4, 2022}{New York, NY, USA}
\acmBooktitle{3rd ACM International Conference on AI in Finance (ICAIF
'22), November 2--4, 2022, New York, NY, USA}
\acmDOI{10.1145/3533271.3561656}
\acmISBN{978-1-4503-9376-8/22/10}

\RequirePackage[
  datamodel=acmdatamodel,
  style=acmnumeric,
  maxcitenames=1,
  maxbibnames=1,
  ]{biblatex}

\usepackage{graphicx}
\usepackage{url}
\usepackage{caption}
\begin{document}

\title{A Deep Learning Approach for Dynamic Balance Sheet Stress Testing}

\author{Anastasios Petropoulos}
\affiliation{%
  \institution{Cyprus University of Technology}
  \city{Limassol}
  \country{Cyprus}
  \postcode{3036}
}
\email{apetropoulos@bankofgreece.gr}

\author{Vassilis Siakoulis}
\affiliation{%
  \institution{Cyprus University of Technology}
  \city{Limassol}
  \country{Cyprus}
  \postcode{3036}
}

\author{Konstantinos P. Panousis}
\affiliation{%
  \institution{Cyprus University of Technology}
  \city{Limassol}
  \country{Cyprus}
  \postcode{3036}
}

\author{Loukas Papadoulas}
\affiliation{%
 \institution{Ethical AI Novelties Ltd.}
 \city{Limassol}
 \country{Cyprus}}

\author{Sotirios Chatzis}
\affiliation{%
  \institution{Cyprus University of Technology}
  \city{Limassol}
  \country{Cyprus}
  \postcode{3036}
}

\renewcommand{\shortauthors}{Petropoulos, Siakoulis, Panousis et al.}

\begin{abstract}
  In the aftermath of the financial crisis, supervisory authorities have considerably altered the mode of operation of financial stress testing. Despite these efforts, significant concerns and extensive criticism have been raised by market participants regarding the considered unrealistic methodological assumptions and simplifications. Current stress testing methodologies attempt to simulate the risks underlying a financial institution’s balance sheet by using several satellite models. This renders their integration a really challenging task, leading to significant estimation errors. Moreover, advanced statistical techniques that could potentially capture the non-linear nature of adverse shocks are still ignored.  This work aims to address these criticisms and shortcomings by proposing a novel approach based on recent advances in Deep Learning towards a principled method for Dynamic Balance Sheet Stress Testing. Experimental results on a newly collected financial/supervisory dataset, provide strong empirical evidence that our paradigm significantly outperforms traditional approaches; thus, it is capable of more accurately and efficiently simulating real world scenarios.
\end{abstract}



\maketitle

\section{Introduction \& Motivation}

Financial Stability constitutes a core component of economic prosperity for countries and individuals. The recent financial crises have had a significantly adverse impact on the life of many individuals across the globe, being a major cause of significant income reduction, rising unemployment and economic slowdown~\cite{otker2013}. Moreover, it was (and still is) undoubtedly established that the currently employed methods of risk management are greatly inadequate; they fail to provide \textit{early warnings} to central governments and central banks in order to \textit{proactively intervene} and \textit{prevent} such adverse financial events. Banks, regulatory authorities, and international organizations (like IMF) performed stress testing (ST) exercises long before the financial crisis of 2007. Nevertheless, ST exercises before Lehman’s default \textit{failed} to predict the \textit{unprecedented economic turmoil}, because they disregarded the propagation channels of a default event through the whole micro and macro dynamics of the global interconnected financial system. It is apparent that the \textit{non-linear relationships} that were realised between the macro-economy and the financial balance sheets were not sufficiently captured due to the broad use of simple linear regression models. Further weaknesses also present in the validation function of ST frameworks also decreased the confidence in the quantification of the impact of an adverse scenario in the banking system.  Since then,  market participants and regulators have performed rigorous STs by \textit{expanding the scenarios to be assessed}, using \textit{more granular data}, and in some cases \textit{attempting to quantify second round effects stemming from a liquidity shock} or \textit{from the default of a counter-party}. 

As a post-crisis response, supervisory regulation has attempted to mitigate some of the shortfalls of current approaches by collecting significant amounts of granular information towards a more proactive supervision. This constitutes a significant step in the right direction for banking supervision via the creation and analysis of big datasets. However, it is surprising that regulators still refrain from exploring more advanced statistical techniques from other fields, such as Deep Learning (DL). These may have the potential to further facilitate the extraction of information regarding the risks in their banking systems. Such a capacity will in turn allow regulators to spot further, and unnoticed by the current frameworks, weaknesses in the supervised financial entities.

Machine learning (ML) algorithms have dramatically improved the capabilities of performing Pattern Recognition, e.g.,  automatic speech recognition, computer vision and forecasting, offering state-of-the-art performance in various scientific fields like language modeling and biology. They are empirically proven to effectively deal with high dimensional data and their structure allows for employment in streaming sequences using continuous learning algorithms, recognizing new and evolving patterns in time-series data. Recent studies suggest that ML techniques could lead to better predictive performance in financial time series modelling problems~\cite{chatzis2018a, fischer2018a, kraus2017a}. This may yield improvements in their performance over time, offering increased capabilities to capture non-linear relationships, and decompose the noise that often exist in financial data. They can easily cope with modeling multivariate time series, therefore capturing the full spectrum of information contained in financial datasets.

Motivated by the recent trends in the ML literature, this empirical study introduces a novel paradigm for ST using DL algorithms to model banks' financial data in a holistic way. In particular, shocks are propagated to each banks' balance sheets by simultaneously training deep neural networks with macro and financial variables. Thus, we we take advantage of their capabilities to capture more information hidden in big datasets. We develop inference algorithms for our networks, suitable for learning financial time series data on a multivariate forecasting setup.

The contributions of this study are two-fold:
\begin{enumerate}
    \item We present a collected dataset regarding the United States Banking System from the database of the Federal Deposit Insurance Corporation (FDIC). The dataset covers a 9-year period with quarterly information with more than $175,000$ records. It comprises an extended set of variables that fully describe the financial status of each bank along with information concerning common macro-economic variables. This will allow for capturing  underlying dependencies between micro and macro variables. 
    \item We propose a holistic framework for balance sheet ST, which overcomes the limitations of current approaches, yielding more robust and close to reality results by loosening the static balance sheet assumption. Our analysis lies at the intersection of computational finance and statistical ML, leveraging the unique properties and capabilities of deep neural networks towards increasing the prediction efficacy. Under this framework, forecasting of balance sheet items can be heavily supported by DL, to better simulate the propagation channels of the macro economy into the financial institutions business models.  Our vision is to provide a ST framework that can serve as an early warning system for financial shocks on individual banks' balance sheets. Thus, we develop networks that \textit{allow for some insights on the reason why a trained model generates some prediction}, facilitating further investigation and easier adoption by regulatory bodies for real-world employment. This is in contrast to existing deep networks for time-series data, such as Transformers\cite{vaswani} and LSTMs\cite{HochSchm97}.
\end{enumerate}

This study is organized as follows. In Section 2, we focus on the related literature review on financial institutions Stress Testing. Section 3 describes the data collection and processing. In Section 4, we provide details regarding the estimation process of the various ST frameworks examined in this study. In Section 5, we compare across methodologies and provide experimental results using a separate test dataset of financial balance sheet sequences of data. Finally, in Section 6, we summarize the performance of the proposed methodology, we identify any potential weaknesses and limitations, while also discussing areas for future research.   

\begin{figure}[!t]
	\centering
	\includegraphics[clip, trim=2.8cm 2cm 2.8cm 2cm,scale= 0.5]{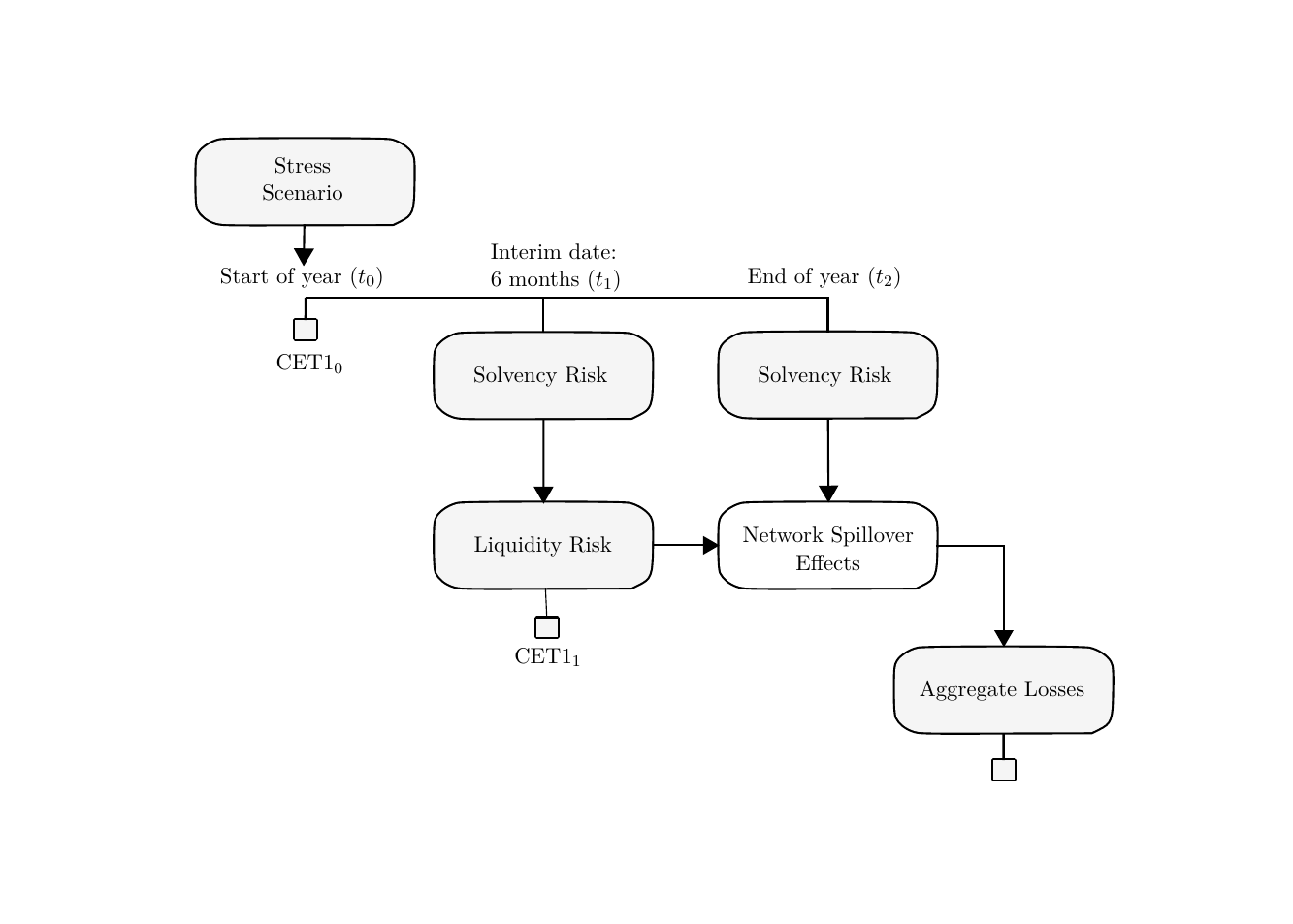}
	\caption{Feedforward architecture of currently established stress testing frameworks.}
	\label{fig:current_ff}
\end{figure}

\section{Related Background}

A Stress Testing engine typically comprises four distinct elements: (i) the perimeter of risks subjected to stress, (ii) the scenario design, (iii) the calculation engine that transforms the shocks into an outcome in Banks' balance sheet, and (iv) a measure of the outcome~\cite{borio2012}. The most famous publicly available ST exercises are: EBA~\cite{eba2018}, CCAR (FED)~\cite{fed}, PRA - Bank of England~\cite{Angleterre2013}, ECB (top down)~\cite{stampe2017,henry2013}, Bank of Canada~\cite{anand2014a}, Central Bank of Austria (ARNIE)~\cite{feldkircher2013a}, IMF~\cite{hasan2011a}, Bank of Greece (Diagnostic Exercise)~\cite{greek}.

To estimate the impact of an adverse shock in the economy, all the aforementioned exercises follow a left-to-right flow. One key aspect is their time horizon: they estimate future losses for banks, ranging from 2 to 5 years. During this period, the macro-economic scenarios are provided. In turn, these are then passed through each financial institution to project their \textit{P}\&\textit{L} and \textit{Risk Weighted Assets} (RWA),  and eventually estimate \textit{capital} using regulatory hurdle rates. A few exercises incorporate a second-round effects mechanism for the banking system to account for any potential contagion risk. However, macroeconomic feedback effects, e.g., the impact of a significant institution becoming insolvent in the macro economy, are usually not considered in such frameworks. ST performed under this rationale, can mainly serve as a tool to challenge the recovery plans of banks and to assess their viability; yet, their role as an early warning system is questionable. 

As Drehmann~\cite{drehmann2011a} aptly points, systemic banking crises are reflected in \textit{credit performance} and \textit{property prices}, and usually appear at the high-point of the medium-term financial cycle. Therefore, crises start before being depicted in macro scenarios. According to Borio~\cite{borio2012}, a system is not fragile when a large financial shock materializes. Instead, it is when even a small negative change in financial and macro variables is amplified through the different dynamic system relationships and can lead to a systemic shock. For example, after the default of Lehman, the financial market crashed and the US GDP exhibited a sharp decrease causing a structural break in the macro data time series. 

Current versions of ST assume a \textit{static over time} macro scenario, without modeling or tracking the path dependent nature of the multi-step decision process and financial behaviour that actually takes place from all economic participants~\cite{bookstaber2014b}. Further, it is well understood that the currently employed statistical techniques fail to adequately model the underlying non-linear relations. With such dynamics, risks under the current globalized market tend to be amplified when a stress event occurs, leading to a chain of events unpredictable from the static nature of typical stress tests. Under \textit{stressed conditions}, the \textit{relationships between modeled variables are non-linear}~\cite{drehmann2007a,juselius2011a} and \textit{exhibit structural breaks}~\cite{alfaro2009a}. 

Apart from explicitly not modeling non-linear dynamics, another reason why ST frameworks fail to capture such relationships is that they are composed by \textit{individual} models, usually combined in a \textit{subjective}, \textit{qualitatively} manner. On this basis, a small single-step prediction error in early stages, could accumulate and propagate when combined without taking the correlation of the financial variables into consideration. This often leads to poor performance and prediction accuracy. Not only that, but such standalone models can lead to double counting effects or overestimating the impact stemming from changes in the predefined macro variables. Finally, as one would expect,  uni-variate setups are not able to appropriately model complex correlated distributed variables exhibiting non-linear behaviors.

Unfortunately, the reliability of the final estimation of ST suffers from yet another modeling decision; that is, the usually employed modeling simplifications. Consider for example the EBA EU wide ST; this constitutes a bottom-up exercise covering only specific risk on banks individuals balance sheet based on a macro scenario that commonly employs such simplifications. Specifically, one of the weaknesses in EBA methodology is the \textit{static balance sheet assumption}, i.e., assets and liabilities remain constant over the horizon, ignoring potential management actions and new generation of loans. In addition, mitigation actions are often taken into account after ST is finalized and through a strong qualitative overlay and not in a dynamic way~\cite{eba2018}.

On the other hand, System-wide ST exercises on micro-prudential level, heavily rely upon interaction of individual banks and with respect to data analytics, propagating the macro scenarios to their balance sheet. Thus, estimation is not performed in a unified statistical process. In contrast, it inherits the model deficiencies and forecast errors embedded in each banks' individual models. The heterogeneity in the results significantly increases estimation errors and there is no robust process for regulators to account for it. Thus, the need for independent central modeling for simulating the financial system is of great significance~\cite{hirtle2015}. This mode of operation is further rendered imperative, considering that the ST process involves the disclosure of the methodological framework to all market participants. In this context, there exist evidence of second round effects regarding the accounting treatment of banks. Specifically, based on a recent study~\cite{gounopoulos2016a}, banks participating in regulatory exercises tend to manipulate their provisions for credit risk to absorb the impact of the upcoming ST. 

ST outcomes in current regulatory exercises heavily depend on \textit{regulatory ratios}, e.g., \textit{Capital Adequacy Ratio}, which in turn are \textit{highly dependable on the estimation of RWA}. Empirical evidence suggest that relying on the internally considered (by the financial institutions) risk weights under the Basel Framework, can lead to \textit{underestimation in capital needs}~\cite{bis}. This stems from the \textit{significant variability} of different internal models of banks. Moreover, the employed regulatory framework for assessing the RWA, \textit{cannot capture the hidden risk in banks complex portfolio structure}. In recent literature~\cite{ferri-a}, there is evidence that  more sophisticated banks in particular (A-IRB),  may perform regulatory arbitrage and manipulate their true risks to lower their capital requirements. Thus, robust macro modelling of the RWA using an independent top-down model is of the utmost importance to account for such cases. 

Although significance progress in designing ST has been made, there are still concerns that this type of exercises cannot be used as early warning systems for financial distress~\cite{otker2013}. We have already outlined the series of weaknesses and inefficiencies regarding ST exercises performed by either regulators or individuals banks.

In this work, we aim to address the various deficits of current frameworks by introducing a novel modeling paradigm, dubbed \textit{Deep Stress}. The core innovation of our proposed approach is the amalgamation of \textit{advanced statistical techniques}, i.e., Deep Neural Networks, and \textit{meticulous analysis and incorporation of financial and economic variables} towards \textit{dynamic balance sheet stress testing}. 

We identify the main channels of risk propagation in a recurrent form to account for all existing evidence of feedback effects in a financial institutions’ balance sheet. Current architectures are constrained by the use classical econometric techniques which offer limited capabilities for simulating complex systems. Our approach accounts for non-linear and temporal patterns in banks’ balance sheets, providing a dynamic modeling approach. We achieve this by taking account the dynamic nature of banks' metrics and the whole structure of each bank’s balance sheet. We bypass the commonly employed uni-variate modeling and combination restriction since DNNs are able to capture the cross-correlation between balance sheet items and the macro economy.  The non-linear relationships that materialize under adverse macroeconomic conditions can be more efficiently be captured due to the structure and capacity of DNNs.  Our proposed framework envisages the effective capturing of such underlying dynamics inherent in a financial distress. Simultaneously, it will allow for determining the amplification channels leading to structural breaks.

Our modeling approach strikes a balance between capturing the determinants that strongly affect the health of a financial institution, while at the same time, developing a dynamic balance sheet simulator engine for establishing an early warning system to predict bank failures under an adverse scenario. The modeling framework that we implement, captures temporal dependencies in a bank’s financial indicators and the macro economy. 

We apply our methodology on a newly collected dataset (from publicly available data) that we describe next. All models are developed in a uniform manner, thus making the process of validation and error correction more feasible to be performed centrally.

\section{Data Collection and Processing}

The dataset introduced in this is study, concerns the United States banking system. Specifically, we have collected information on \textit{non-failed}, \textit{failed} and \textit{assisted entities} from the database of the Federal Deposit Insurance Corporation (FDIC); an independent agency created by the US Congress in order to maintain the stability and the public confidence in the financial system. The collected information is related to all US banks, while the adopted definition of a default event in this dataset includes all bank failures and assistance transactions of all FDIC-insured institutions. Under the proposed framework, \textit{each entity} is categorized either as \textit{solvent} or as \textit{insolvent} based on the indicators provided by FDIC. Observations referring to failed banks are excluded from the analysis since ST is performed on healthy financial entities. 

\begin{figure}
	\centering
	\includegraphics[width=\linewidth]{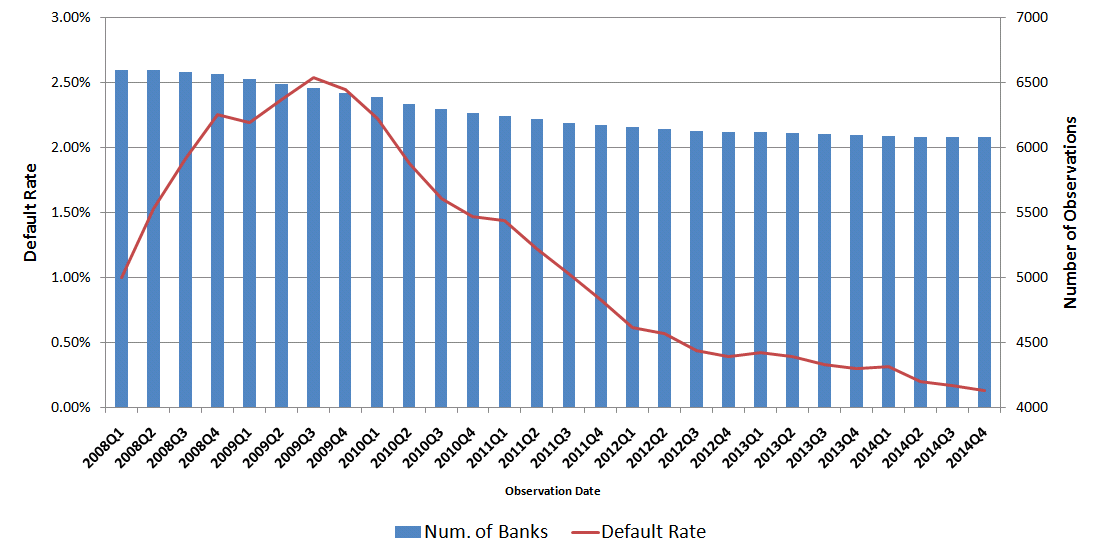}
	\caption{US Financial Institutions: Historical overview of the 2008-2014 period of failed entities (source: FDIC)}
	\label{fig:usa_historical}
\end{figure}

The dataset covers $2007-2015$: a $9$-year period with quarterly information resulting in more than $175,000$ records. The selected time period, approximates a full economic cycle, in terms of the Default Rate evolution. Fig. \ref{fig:usa_historical}, shows the number of records included in each observation quarter and the corresponding default rate. From a supervisory perspective, most of the financial institutions in the sample, apply the standardized approach for measuring the Credit risk weights assets based on the United States adaptation of the Basel regulatory framework~\cite{castro2017a}. 

The dataset was split into three parts: (i) an \textit{in-sample training} set, comprising data pertaining to the $80\%$ of the examined banks over the observation period $2008-2013$, and amounting to $101,641$ observations, (ii) an \textit{out-of-sample validation set} that facilitates hyper-parameter tuning of deep neural networks that allows for increased generalization capabilities, and included the remaining $20\%$ of the observations for the period $2008-2013$, amounting to $25,252$ observations, and (iii) an \textit{out-of-time test} set that spans over the $2014-2015$ observations period, comprising $48,756$ records that is used for the final performance evaluation.

In all splits, the dependent (target) variable is the \textit{Capital Adequacy Ratio} (CAR) of each bank in the end of the one year forecast horizon. We perform model fitting using \textit{exclusively the constructed training sample}. For model selection, we employed five-fold cross-validation, using predictive accuracy as our model selection criterion, i.e.,  the CAR prediction error. Performance evaluation results are finally assessed on the available \textit{test set}, to allow for evaluating the generalization capacity of the developed models. 

To capture the complex relationships between banks and macro-economic metrics, we consider an extended set of variables that comprise: (i) variables that fully described the financial status of each bank in the samples, and (ii) quarterly observations of the most commonly used macro-economic variables. Macro variables are the main input in the models developed, since they are important for scenario analysis under an ST framework. The current model setup includes contemporaneous macro variables along with 3 year lags. The intuition for this approach is to build models for \textit{scenario prediction} which is the main methodology for ST modeling.

Specifically, the macro variables are (i) \textbf{GDP}: Gross Domestic Product growth, (ii) \textbf{EXPORT}: US Total Exports growth, (iii) \textbf{GOVCREDIT}: Government Credit to GDP,
(iv) \textbf{DEBT}: US public debt to GDP, (v) \textbf{GOVEXP}: US government expenditure to GDP, (vi) \textbf{INFLAT}: US inflation,
(vii) \textbf{RRE}: House Price Index growth, (viii) \textbf{UNR}: Unemployment Rate, 
(ix) \textbf{YIELD10Y}: 10Y US sovereign bonds yields, and (x) \textbf{STOCKS}: US Stock index – S\&P 500 returns.

The relevant stress financial variables for simulating the profitability and the risk weighted assets of each financial institution are:  (i) \textbf{NLOAN}: Net loans exposure,
(ii) \textbf{DEP}: Total Deposits,
(iii) \textbf{DDEP}: Total domestic deposits,
(iv) \textbf{ASSET}: Average Total Assets,
(v) \textbf{EASSET}: Average Total Earning Assets,
(vi) \textbf{EQUITY}: Average Total Equity,
(vii) \textbf{LOAN}: Average total loans,
(viii) \textbf{CFD}: Deposits Cost of funding, 
(ix) \textbf{YEA}: Yield on earning assets,
(x) \textbf{NFIA}: Noninterest income to average assets,
(xi) \textbf{RW}: Risk Weight Density, 
(xii) \textbf{LOSS}\_\textbf{LOAN}: Loss allowance to loans,
(xiii) \textbf{RWA}: Total risk weighted assets, and
(xiv) \textbf{CAR}$\%$: Total Risk-based Capital Ratio.

Modeling for the evolution of the balance sheet is performed on the growth rate of 4 key financial items: \textit{Deposits}, \textit{Total Earning Assets}, \textit{Total Loans} and \textit{Total Assets}. 
In order to capture the idiosyncratic characteristics of each financial entity, 3-year lags are included in the training process for each financial variable. In the final model setup, the use of multiple years financial and macroeconomic variables allows for capturing internal trends of key items of a bank' balance sheet and also the degree each entity is affected by the status of the US economy.

\section{Model Development}

 To investigate the capabilities and efficacy of the proposed framework for ST, we further implement and compare our results to two real-world methods for balance sheet forecasting. Specifically, we develop (i) a \textit{Constant Balance Sheet} approach, following the framework adopted by EBA to perform EU wide ST~\cite{eba2018} and (ii) a \textit{dynamic balance sheet approach}, supported by a \textit{group of satellite models to forecast individual financial variables} used by other regulatory authorities like ECB for macro prudential ST. Next, we provide an overview of the overall setup of the study and the technical details of the three individual implementing methods.

\subsection{General Setup of the Study}

The main aim of a micro prudential solvency ST framework is the projection of a financial institution's CAR or the more recent Core Equity Tier (CET)-1  ratio. To this end, we develop a DNN structure which receives as input the Macro variables and Balance sheet components described in the previous section. The produced output corresponds to the balance sheet and profitability structure of the bank on one year horizon, as measured by $9$ \textit{core} variables, namely: \textit{Net loans}, \textit{Deposits}, \textit{Assets}, \textit{Earning Assets}, \textit{Cost of funding}, \textit{Yield on earning assets}, \textit{Noninterest income to assets}, \textit{Risk Weight Density} and \textit{Cost of Risk} (loss allowance to loans).
	
	We focus on forecasting of CAR, since CET-1 ratio was introduced under Basel III and is not available throughout our dataset. Our aim is to project the in a one-year-ahead the CAR ratio of each financial institution in the sample. CAR, by definition, is the ratio of a \textit{bank’s Capital} over the \textit{Risk Weighted Assets} at each time point $t$. In order to simulate the core mechanics of an ST framework, we simulate the evolution of the key financial variables of a financial institution's balance sheet. Specifically, we project one-year ahead, the \textit{evolution of the capital} and the \textit{risk weighted assets} to forecast the one-year ahead CAR. The approach followed to adjust the capital at time $t$ reads: 
	\begin{align*}
	\begin{split}
	\text{Capital}_t = &\text{Earnings from Assets}_t - \text{Loans Loss Provisions}_t \\&
	+ \text{Net Fees and Commissions}_t \\
	&- \text{Cost of Funding from Deposits}_t + \text{Capital}_{t-1}
	\end{split}
	\end{align*}
	To adjust the capital of each entity, we model 8 key financial variables. The first four concern the dynamic evolution of a balance sheet, i.e., the growth of the asset and liability side: the \textit{growth rate} of \textit{Deposits}, \textit{Total loans}, \textit{Total Assets}, \textit{Total Earning Assets}. The remaining 4, the yield in the next year of each item from the asset or liability side: \textit{cost of risk of loans}, \textit{yield on earning assets}, \textit{yield on deposits} and \textit{yield of net fees and commissions of total assets}.  
	
	The RWA are adjusted in 3 different ways depending on the ST methodology: (i) for DL, we project the growth of the RWA, (ii) for satellite modelling, a dedicated model is trained to project the RW density of each financial institution in the sample, while (iii) for the constant balance sheet approach, we assume RWA remain constant for one year.

	\subsection{Constant Balance Sheet Modelling Setup}
	
	In this approach, all balance sheet items are assumed \textit{constant, along with the RWA metric for one year}. To project yields of assets and liabilities, we \textit{combine} the respective uni-variate satellite BMA models,  while assuming \textit{zero growth in the balance sheet} in order to \textit{project the CAR ratio one year ahead}. 
	
	\subsection{Satellite Modeling - Bayesian Model Averaging}
	
	Satellite models are used for \textit{uni-variate estimation of the impact} of \textit{standalone} balance sheet items in current ST frameworks~\cite{stampe2017}. A commonly employed technique by both regulators and the banking industry is \textit{Bayesian Model Averaging} (BMA).  The main intuition for its employment is to account for the \textit{uncertainty} surrounding the core determinants of risk dynamics, especially in a period of recession. This way, short time-series of balance sheet realizations for ST  can be handled. BMA offers the ability to perform multivariate modeling, including all potential predictors with different weights, while the output of each trained model remains uni-variate. 
	
	In BMA, a pool of equations is generated using a randomly selected subgroup of determinants, and an appropriate \textit{weight} is assigned to each model, reflecting its \textit{relative forecasting performance}. The posterior model probability is then formed by aggregating all equations using their corresponding weights. In the first step, the number of equations estimated is large enough to \textit{capture all possible combinations of a predetermined number of independent variables}. BMA addresses\textit{ model uncertainty} and \textit{mis-specification} via selected explanatory variables in a simple linear regression problem. 
	
	Let $Y_t$ be the dependent variable and $X_t$ the explanatory variables. Then, assuming a linear model structure with error $\epsilon_t$:
	\begin{align*}
	Y_t &= \alpha_\gamma + \beta_\gamma X_{\gamma, t} + \epsilon_t, \qquad
	\epsilon_t \sim \mathcal{N}(0, \sigma^2 I)
	\end{align*}
	where $\alpha$ is a constant and $\beta$ are the regression coefficients.
	When several potential explanatory variables are present, selecting a correct combinations becomes quite complex.  A simple linear model that includes all variables is inefficient or even infeasible with a limited number of observations. It can result to \textit{overfitting}, \textit{multi-collinearity} and\textit{ increased necessity for manual re-estimations to account for non-significant determinants}. BMA tackles the problem by estimating models for all possible combinations of $\{X\}$ and constructing \textit{a weighted average} over all of them.
	
	Under the assumption that $X$ contains $K$ potential explanatory variables, BMA estimates $2^K$ combinations, and thus, $2^K$ models. Applying Bayes' Theorem, model averaging is based on the posterior model probabilities:
	\begin{align*}
	p(M_\gamma \cup Y, X) &= \frac{p(Y \cup M_\gamma, X) p(M_\gamma)}{p(Y \cup X)} = \frac{p(Y \cup M_\gamma, X) p(M_\gamma)}{\sum_{s=1}^{2^K} p(Y \cup M_s, X) p(M_s)}
	\end{align*}
	The denominator is common in all models; thus, the posterior model probability is proportional to $p(Y \cup M, X)$ which reflects the probability of the data given the model $M$. Thus, the corresponding weight assigned to each model is measured by using $p(M_\gamma \cup Y, X)$. $p(M)$ denotes the prior belief of how probable model $M$ is before analyzing the data. We can then  infer the model's weighted posterior distribution for the coefficients $\beta$, yielding:
	\begin{align*}
	p(\beta \cup Y, X) = \sum_{\gamma=1}^{2^K} p(\beta \cup M_\gamma, Y, X) p(M_\gamma \cup X, Y)
	\end{align*}
	We assume a uniform prior for each model. Regarding the marginal likelihoods and the posterior distributions, we use the g prior \cite{zellner1986onassessing}, while the prior for the coefficients is assumed to be a normal distribution with pre-specified mean and variance.

	\subsection{Deep Learning}
	
	 Deep learning is a highly active field of research, having recently achieved significant breakthroughs in the fields of computer vision and language understanding. They have been extremely successful in diverse time-series modeling tasks, including machine translation~\cite{vaswani} and recommendation engines~\cite{quadrana2017a}. However, their application in the field of finance is rather limited. To the best of knowledge, our work constitutes one of the first in the literature that considers DL to address the challenging financial modeling task of \textit{dynamic financial balance sheet stress testing}. 
	 
	 Let us consider input data $ X \in \mathbb{R}^{N\times D}$, containing $N$ observations, with $D$ features each. In a traditional DNN hidden layer, we compute an inner product between the input and a weight matrix $W \in \mathbb{R}^{D\times K}$. The resulting \textit{activation} is then passed through a non-linear function $\sigma(\cdot)$, yielding the final output $\boldsymbol Y\in\mathbb{R}^{N \times K}$ of each layer:
	\begin{align}
	y_{nk} = \sigma \left(\sum_{d=1}^d w_{dk} x_{nd} \right)
	\end{align}
	Each DNN comprises multiple such layers, connected in an hierarchical manner. The most widely used non-linear activation function is the Rectified Linear Unit (ReLU): $
	relu(x) = \max (0, x)$.
	
	In this work, we additionally consider a radically different paradigm of latent unit operation, based on the biologically-inspired Local Winner-Takes-All (LWTA) mechanism. In this context, hidden units compete for their outputs; the winner gets to pass its output to the next layer, while the rest are zeroed out. LWTA activations have been shown to exhibit significant properties such as \textit{noise suppression}, \textit{adversarial robustness}, \textit{compression capabilities} and facilitate learning of \textit{diversified representations}\cite{panousis2019nonparametric, panousis21a, panousis21bdl, panousis2022competing}. Thus, we explore their potency in the financial domain.
	
	In our setup, multivariate DL networks will learn the balance sheet of financial institutions, and separately generate yearly forecasts after receiving historical values of banks' previous economic states. The hierarchical transmission of observed data between cascading layers of abstraction can decompose the structure of a bank balance sheet and foster the multivariate representation of the financial variables for better capturing the correlations between various assets and liabilities. Thus, we can simultaneously model the balance sheet as a whole instead of using satellite models of regular ST frameworks. DNNs further facilitate dynamic balance sheet projection through their non-linear nature, offering a more realistic approach for ST. Information flows through the system as a vector of macro and financial variables describing the state of both the bank and the macro economy at any time stamp during the forecast period. The input vector contains around 60 variables and the output 9 variables.  The considered DNN is capable to model the lead lag relationships between macro, banks', financial, and sovereign variables. Finally, through this multivariate forecasting setup on individual balance sheet, we simultaneously model the RWA evolution of each bank and connect it to the macro environment. 
	
	\subsubsection{Bayesian Deep Learning}
	
	Conventional DNN architectures compute point estimates of the unknown values, i.e., each layer's weights, without taking into consideration any \textit{prior information} and \textit{without any uncertainty estimation} of the produced values. The Bayesian framework offers a flexible and mathematically founded approach to incorporate prior information and uncertainty estimation by explicitly employing \textit{model averaging}. The Bayesian treatment of particular model has been shown to increase its capacity and potential, while offering a natural way to assess the uncertainty of the resulting estimates.  To this end, we additionally assess the performance of such Bayesian Neural Networks (BNNs). Specifically, we impose a prior Normal distribution over network weights, seeking to infer their posterior distribution given the data. Since the marginal likelihood is intractable for the considered architectures, from the existing Bayesian methods, we rely on approximate inference and specifically on Stochastic Gradient Variational Bayes (SGVB)\cite{Kingma2014} and optimize the resulting Evidence Lower Bound (ELBO) expression \cite{Blei_2017}.

	\section{Experimental Evaluation}

	\textbf{BMA}: Before applying BMA, we remove and linearly interpolate the outliers. We employ a unit information prior (UIP), which sets g=N commonly for all estimation models, and rely on a birth/death MCMC algorithm (20000 draws) due to the large number of included covariates. We fix the number of burn-in draws for the MCMC sampler to 10000. A ``random theta’'' prior \cite{ley2008a} is employed, comprising a binomial-beta hyper prior on the a-priori inclusion probability. This prior has the advantage of being less ``tight'' around a-priori expected model size, i.e., the average number of included regressors. Thus, it reflects prior uncertainty about model size more efficiently. For robustness purposes, we also considered  varying the prior using the Fernandez propositions~\cite{fernandez2001a}; however, the results were not substantially different.
	We develop all satellite models using the BMS R package{\footnote{https://cran.r-project.org/web/packages/BMS/index.html}}. After training, 9 BMS models are developed: 4 for the growth of balance sheet items, 4 models are forecasting the yields of a various assets and liabilities and one model for forecasting the RW assets density. 
	
	\textbf{DNNs}: DNNs typically comprise a massive amount of trainable parameters; thus, it is essential to employ appropriate techniques to prevent them from overfitting.  Thus, we consider Dropout \cite{srivastava2014a} with ReLU activations for standard non-bayesian training, using the Apache MXNET toolbox of R. We postulated deep networks that are up to five hidden layers deep, comprising various numbers of neurons. Increasing the number of layers did not result in any significant improvement. Model selection using the validation set was performed by maximizing the RMSE metric on the projected CAR. For BNNs, we use Tensforflow \cite{tensorflow2015-whitepaper}, and develop our models from scratch.
	
	\begin{table}
		\centering
		\caption{Comparison of the predicted one year ahead CAR by ST approach for all banks and only for Large financial institutions (more than 200 billion in assets).}
		\label{table:car}
		\begin{minipage}[b]{\linewidth}
		\centering
		\resizebox{\linewidth}{!}{	
			\begin{tabular}{|c|c|c|}
				\hline
				All banks in the dataset & Out of Sample CAR & In Sample CAR\\\hline
				Satellite Modelling (BMS) & 20.61 & 17.07 \\\hline
				Deep Learning (MXNET) & 18.01 & 17.89 \\\hline
				Deep Learning (Bayesian ReLU)  & 18.80 & 17.83 \\\hline
				Deep Learning (Bayesian LWTA) & $\mathbf{19.23}$ & $\mathbf{18.53}$ \\\hline
				Constant Balance Sheet & 20.03 & 17.49 \\\hline
				Actual & 19.33 & 18.73 \\\hline
		\end{tabular}}
		\end{minipage}
		\begin{minipage}[b]{\linewidth}
		\centering
		\resizebox{\linewidth}{!}{
			
			\begin{tabular}{|c|c|c|}
				\hline
				Large Banks ($>$ 200 bl) & Out of Sample CAR & In Sample CAR \\\hline
				Satellite Modelling (BMS) & 15.07 & 11.04 \\\hline
				Deep Learning (MXNET) & 12.7 & 11.12 \\\hline
				Deep Learning (Bayesian ReLU) & $13.2$ & $11.72$\\\hline
				Deep Learning (Bayesian LWTA) & $\mathbf{13.43}$ & $\mathbf{12.13}$ \\\hline
				Constant Balance Sheet & 15.11 & 11.48 \\\hline
				Actual & 13.75 & 14.16 \\\hline
			\end{tabular}
		}
	\end{minipage}
		
	\end{table}
	
	\subsection{Evaluation}
	
	No thorough and consistent framework exists for validating the results of an ST exercise, since the adverse scenario used in their design never materializes. Thus, the success of the ST exercises after the financial crises maybe be circumstantial~\cite{hirtle2015}, since no robust methods are applied to quantify their estimation error. Back-testing methods are important to recognize modeling inefficiencies and fine-tune the estimations, considering specificities in the time-series data that were not captured in the initial calibration and development. To improve the quality of ST, rigorous validation procedures of actual vs predicted financial variables are essential. 
	
	\begin{figure}[t!]
		\centering
		\includegraphics[width=\linewidth]{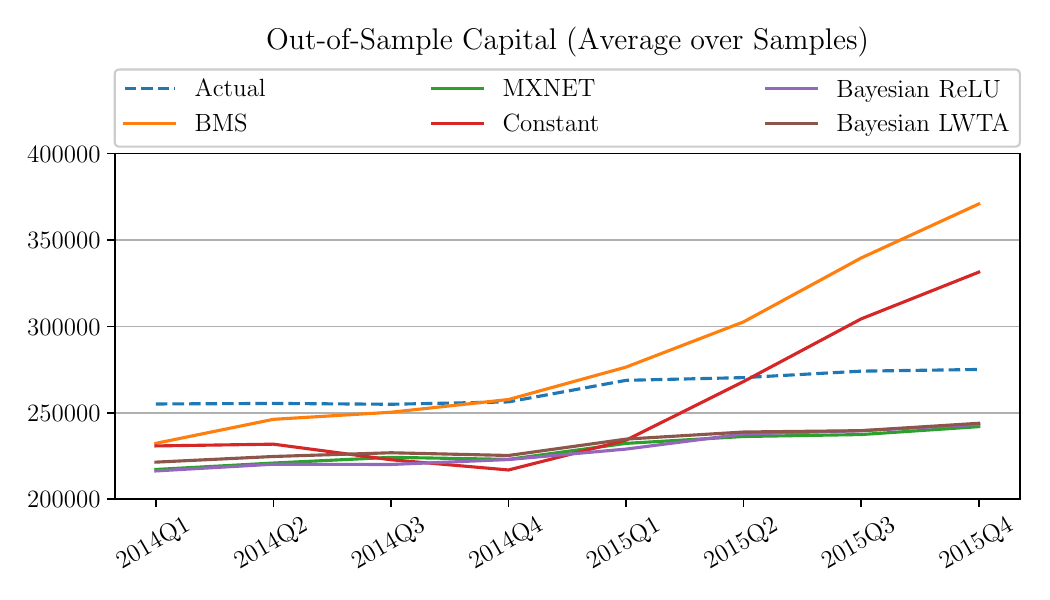}
		\caption{Out-of-sample results of the predicted Capital compared to the actual figures (Whole Sample). }
		\label{fig:capital_average}
	\end{figure}
	
	\begin{figure}[b!]
		\centering
		\includegraphics[width=\linewidth]{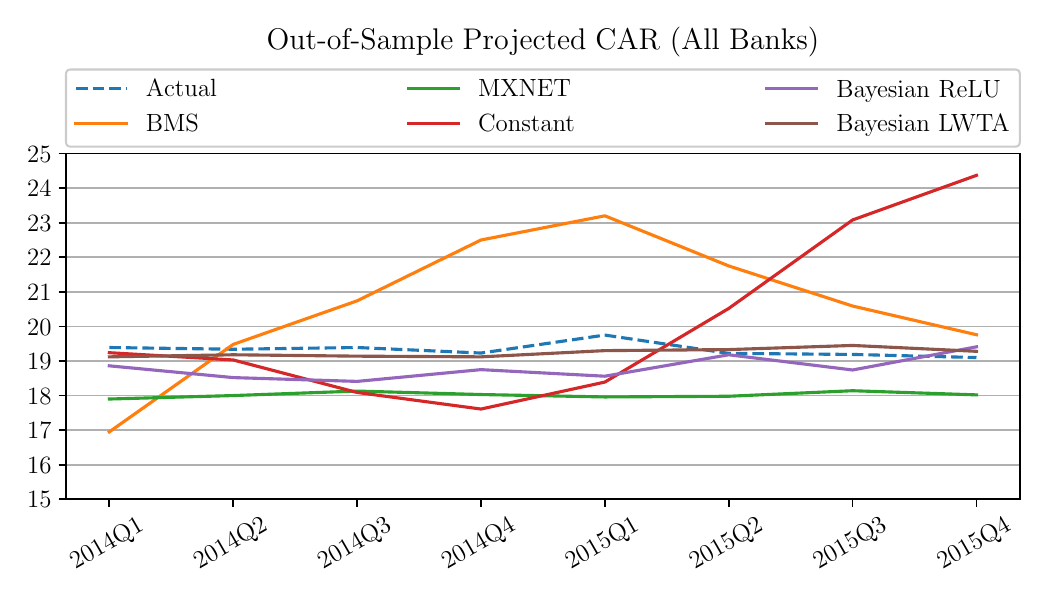}
		\caption{Out-of-sample results of the predicted CAR compared to the actual values (Whole Sample) }
		\label{fig:car_average}
	\end{figure}

	In this study, we perform a thorough validation procedure in order to assess the robustness of our approach. We only use the “in-sample” set, spanning from 2010 to 2013 (16 quarters), to develop out ST frameworks. Then, we assess its performance under the ``Out-of-time'' set; each model is evaluated over a two year (8 quarters) out-of-sample time-period spanning from 2014 to 2015 to assess their generalization capacity. Validation is performed with respect the one year ahead forecast of the CAR ratio. Note that the last two years of the dataset \textit{were not used for model development}.
	Prediction accuracy of the CAR ratio is the main criterion to assess the efficacy of each method. We employ the usual forecast metrics of \textit{Root Mean Square Error} (RMSE), \textit{Mean Absolute Error} (MAE) and the \textit{Mean Absolute Percentage Error} (MAPE). 
	
	\begin{figure}[t!]
		\centering
		\includegraphics[width=\linewidth]{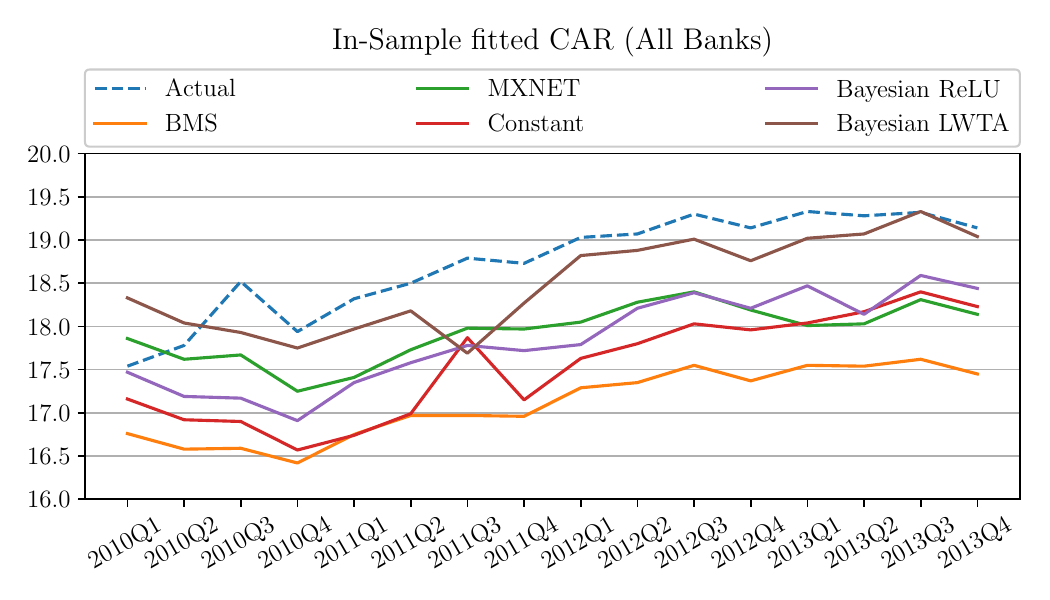}
		\caption{In-sample results of the predicted CAR compared to the actual values (Whole Sample).}
		\label{fig:car_all_banks_in_sample}
	\end{figure}

	In Table \ref{table:car}, the comparative results are depicted. We observe that DL-based algorithms result in the best empirical fit both in-sample and out-of sample sets. Specifically, the predicted average CAR is closer to the actual value, when compared to the commonly employed approaches. Such performance hints to  a more efficient and holistic way to simulate the CAR under a specific set of macro scenarios of key macroeconomic variables using DNNs.  Turning to the specific validation metrics, these are shown in Table \ref{table:car_metrics}. Once again DNNs provide more accurate estimation of the CAR ratio, exhibiting a significant decrease in the forecasting error. It it noteworthy that, by moving from DNNs to BNNs, and when using LWTA activations in particular, we are able to infer richer and subtler dynamics from the data. This translates to increased capacity in modeling nonlinearities and cross-correlations among balance sheet P\&L items. This is also evident from Figs. \ref{fig:capital_average} and \ref{fig:car_average}, where the out-of-sample performance of constant balance sheet and satellite modelling diverge significantly from the actual evolution of Regulatory Capital (Fig. \ref{fig:capital_average}) and the CAR (Fig. \ref{fig:car_average}), despite exhibiting adequate fit for the in-sample set (Fig. \ref{fig:car_all_banks_in_sample}). In stark contrast, average CARs estimated using  DL-based methods appropriately captured the dynamics in the projection period.

	\begin{figure}[b!]
		\centering
		\includegraphics[width=\linewidth]{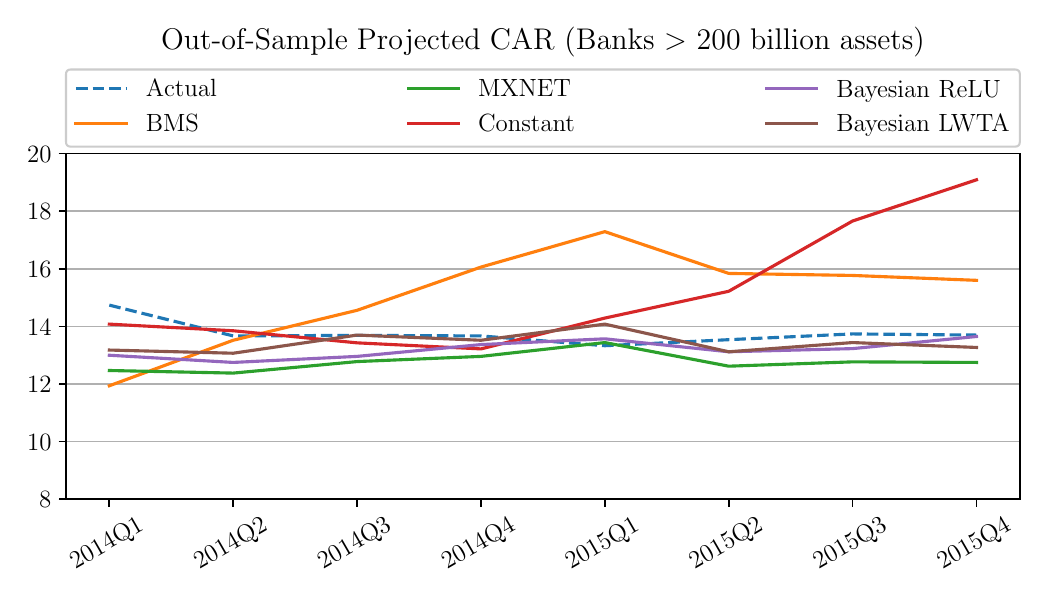}
		\caption{Out-of-sample results of the predicted CAR compared to the actual values (Large Banks in the out-of-sample). }
		\label{fig:car_large_banks}
	\end{figure}
	
	\begin{table}
		\caption{Comparison of the predicted one year ahead CAR by ST approach for all banks and only for Large financial institutions (more than 200billions in assets)}
		\label{table:car_metrics}
		\centering
		\begin{minipage}[b]{\linewidth}
		\centering
		\resizebox{\linewidth}{!}{
			\begin{tabular}{|c|c|c|c|}
				\hline
				\textbf{All banks} & \multicolumn{3}{c|}{Out of Sample (2014Q1-2015Q4)} \\\hline
				& \textbf{RMSE} & \textbf{MAPE} & \textbf{MAE} \\\hline
				Satellite Modelling (BMS) & 11.32 & 2.88 & 0.15 \\\hline
				Deep Learning (MXNET) & 11.18 & 2.36 & 0.12 \\\hline
				Deep Learning (Bayesian ReLU)  & 15.36 & 2.12 & 0.10 \\\hline
				Deep Learning (Bayesian LWTA) & $\mathbf{10.75}$ & $\mathbf{1.77}$ & $\mathbf{0.09}$\\\hline
				Constant Balance Sheet & 11.15 & 2.85 & 0.15 \\\hline
				& \multicolumn{3}{c|}{In Sample (2010Q1-2013Q4)}\\\hline
				Satellite Modelling (BMS) & 13.46 & 2.58 & 0.16 \\\hline
				Deep Learning (MXNET) & 13.49 & 2.55 & 0.15 \\\hline
				Deep Learning (Bayesian ReLU)  & 16.58 & 2.41 & 0.15 \\\hline
				Deep Learning (Bayesian LWTA) & 18.70 & 2.16 & 0.14\\\hline
				Constant Balance Sheet & 17.25 & 2.56 & 0.15 \\\hline
			\end{tabular}
		}
	\end{minipage}
	\begin{minipage}[b]{\linewidth}
	\centering
		\resizebox{\linewidth}{!}{
			\begin{tabular}{|c|c|c|c|}
				\hline
				\textbf{Large Bank}s ($>$ \textbf{200 bl}) & \multicolumn{3}{c|}{Out of Sample (2014Q1-2015Q4)}\\\hline
				& \textbf{RMSE} & \textbf{MAPE} & \textbf{MAE} \\\hline
				Satellite Modelling (BMS) & 3.21 & 2.31 & 0.17 \\\hline
				Deep Learning (MXNET) & 2.28 & 1.97 & 0.15 \\\hline
				Deep Learning (Bayesian ReLU)  & $\mathbf{1.96}$ & 1.56 & 0.12 \\\hline
				Deep Learning (Bayesian LWTA) & 2.04 & $\mathbf{1.51}$ & $\mathbf{0.11}$\\\hline
				Constant Balance Sheet & 3.56 & 2.58 & 0.19 \\\hline
				& \multicolumn{3}{c|}{In Sample (2010Q1-2013Q4)}\\\hline
				Satellite Modelling (BMS) & 3.44 & 3.14 & 0.23 \\\hline
				Deep Learning (MXNET) & 3.46 & 3.13 & 0.22 \\\hline
				Deep Learning (Bayesian ReLU)  & 3.07 & 2.78 & 0.20 \\\hline
				Deep Learning (Bayesian LWTA) & \textbf{2.76} & \textbf{2.4}2 & \textbf{0.18}\\\hline
				Constant Balance Sheet & 3.27 & 2.94 & 0.21 \\\hline
			\end{tabular}
		}
	\end{minipage}
		
	\end{table}
	
	To further investigate the performance of \textit{DeepStress}, we further focus on a subset of \textit{large financial institutions}, where performance of a robust ST methodology is more important due to their social-economic impact. For the purpose of this investigation, large financial institutions are defined as entities with more than 200 billion in assets. In Tables \ref{table:car} and \ref{table:car_metrics}, we observe that the superiority of DNNs is further affirmed in all error metrics, exhibiting significant drops in the forecasting error in the test sample.  It is worth noting that, even though satellite univariate modelling in the sample dataset was expected to provide a better fitting against the DNN, this is not the case. The considered DNNs are trained on a multivariate setup, attempting to model 9 variables at the same time, and still exhibit better in-sample error against the other two methods. The same pattern also holds in Fig. \ref{fig:car_large_banks}, where the projected CAR is depicted only for the large banks (more than 200 billion in assets).
	
	The experimental results vouch for the efficacy of the proposed paradigm. It is evident that DNNs exhibit higher predicting power compared to all benchmark approaches. In contrast, the commonly considered constant balance assumption, although easier to implement, exhibits the highest error. It is therefore crucial for supervisory authorities to rethink current ST exercises that are based on the constant balance sheet assumption and move towards an advanced dynamic balance sheet approach.

	\section{Conclusions and Future Work}
	
	In this work, we proposed a novel modeling paradigm for regulatory stress testing exercises, namely \textit{DeepStress}. Our main innovation relative to the forecasting economic and financial crisis events literature is that we explore new deep learning-based statistical techniques to tackle the problem of dynamic balance stress testing. DNNs  (and BNNs) were utilized to provide a holistic modeling approach for a bank’s key financial items and dominant macro variables. We performed thorough testing and validation of the proposed approach and compared its performance against two broadly accepted and employed stress testing frameworks: constant balance sheet and satellite dynamic modeling. Our experimental results provide strong empirical evidence for its efficacy. Our approach consistently outperformed the benchmark methods, exhibiting consistent improvement in the forecasting accuracy with respect to the Capital Adequacy Ratio.  The main driver for this higher forecasting accuracy is the potential to model the balance sheet intercorrelation of P\&L items providing better simulation of the banks one-year-ahead activities. DeepStress offers a better dynamic balance sheet simulator, which is a major component in any stress testing framework. Such simulation capacity will allow better capturing that small macro and financial changes that can be amplified exponentially under a crisis event. 
	
	We initially focused on the banking system, as it constitutes the backbone of the global economy, but our paradigm is scalable to other entities, corporate, insurances and shadow banking. We strongly believe that DeepStress should be explored by regulators and financial institutions in order to produce a new generation of stress testing, increasing monitoring and awareness for possible future financial shocks.

\begin{acks}
      This work was co-funded by the European Regional Development Fund and the Republic of Cyprus through the Research and Innovation Foundation (Project: POST-DOC/0718/196).
\end{acks}

\printbibliography

\end{document}